# Windowless Observation of Evaporation-Induced Coarsening of Au-Pt Nanoparticles in Polymer Nanoreactors

*Jingshan S. Du,[†, §] Peng-Cheng Chen,[†, §] Brian Meckes,[‡, §] Edward J. Kluender,[†, §] Zhuang Xie,[‡, §, ⊥] Vinayak P. Dravid,[*, †, §] Chad A. Mirkin[*, †, ‡, §]*

[†]Department of Materials Science and Engineering, [‡]Department of Chemistry, and [§]International Institute for Nanotechnology, Northwestern University, Evanston, Illinois 60208, United States

ABSTRACT. The interactions between nanoparticles and solvents play a critical role in the formation of complex, metastable nanostructures. However, direct observation of such interactions with high spatial and temporal resolution is challenging with conventional liquid-cell transmission electron microscopy (TEM) experiments. Here, a windowless system consisting of polymer nanoreactors deposited via scanning probe block copolymer lithography (SPBCL) on an amorphous carbon film is used to investigate the coarsening of ultrafine (1–3 nm) Au-Pt bimetallic nanoparticles as a function of solvent evaporation. In such reactors, homogeneous Au-Pt nanoparticles are synthesized from metal ion precursors *in situ* under electron irradiation. The non-uniform evaporation of the thin polymer film not only concentrates the nanoparticles, but also accelerates the coalescence kinetics at the receding polymer edges. Qualitative analysis of the particle forces influencing coalescence suggests that capillary dragging by the polymer edges plays a significant role in accelerating this process. Taken together, this work: 1) provides fundamental insight into the role of solvents in the chemistry and coarsening behavior of nanoparticles during the synthesis of polyelemental nanostructures, 2) provides insight into how particles form via the SPBCL process, and 3) shows how SPBCL-generated domes, instead of liquid cells, can be used to study nanoparticle formation. More generally, it shows why conventional models of particle coarsening, which do not take into account solvent evaporation, cannot be used to describe what is occurring in thin film, liquid-based syntheses of nanostructures.

INTRODUCTION

Understanding the factors that control the structure and morphology of nanoparticles,[1-3] thin films,[4-5] and superlattices[6-8] obtained from solution-mediated processes is critical for generating materials with desired catalytic, optical, and electronic properties.[9-13] On the nanoscale, unusual phenomena can be in play, and metastable structures can be observed that are not predicted by bulk phase diagrams. For example, in the case of particles made from Au and Pt, which should lead to a thermodynamically preferred phase separated state, metastable alloy nanoparticles that presumably result from coalescence processes have been observed.[14] The formation of such structures underscores the fact that the interactions between ultrafine nanoparticles and solvents, and their role in the growth of metastable nanostructures, are not completely understood.[15] Indeed, a complete understanding of the structural evolution of nanoparticles into metastable nanostructures would facilitate their controlled synthesis and implementation as reactive catalysts.[14, 16-18]

At present, the study of ultrafine particle dynamics is limited due to the need for high-resolution imaging techniques capable of direct observation of particles in the solution phase. Recently developed liquid cells provide a fluidic environment for *in situ* nanoscale imaging by (scanning) transmission electron microscopy [(S)TEM].[19-22] Through the use of this technology, substantial insights into nanoparticle growth,[23-24] coarsening,[22, 25-28] migration,[29-31] and assembly[32-35] in fluids have been obtained. In particular, it is clear that liquid cells are a powerful tool for studying fluid–nanoparticle interactions in evaporation-mediated particle assembly.[36-37] However, the interaction of ultrafine (1–3 nm) nanoparticles with each other and/or the solvent has not been carefully investigated. Several challenges have prevented such studies in conventional liquid cell experiments, including: 1) the fast evaporation and low contrast of typical liquids makes visualizing and determining the role of solvents in the particle synthesis process difficult on the one-nanometer scale; and 2) the presence of two encapsulation windows leads to high background due to electron scattering, which limits image contrast and resolution as well as the ability to use correlated analytical techniques.

A windowless thin film system with a non-volatile fluid could overcome these challenges. This type of system requires a relatively stable solvent with slow evaporation rates that can carry nanoparticles in the high vacuum environment inside a TEM.[38-39] Polymer nanoreactors have previously been made by dip-pen nanolithography (DPN) and used to generate complex libraries of polyelemental nanoparticles consisting of metals, oxides, and their hybrid structures through a thermally driven process termed scanning probe block copolymer lithography (SPBCL).[40-44] Importantly, the polymer nanoreactors are stable enough to study nanoparticle coarsening at controlled temperatures using *in situ* STEM.[45]

Herein, we use DPN-deposited polymer nanoreactors and *in situ* STEM to study the interactions between ultrafine Au-Pt nanoparticles and the surrounding thin polymer films. A key to being able to accomplish this goal is the observation that polymer [poly(ethylene oxide)-*block*-poly(2-vinylpyridine), PEO-*b*-P2VP] deposition on amorphous carbon films results in features that spread on the surface to generate thin nanoreactors (< 10 nm over large areas at the reactor edge), which enables *in situ* observation of particle growth pathways. When loaded with gold(III) and platinum(IV) precursors, the electron beam can be used to generate 1–3 nm metastable alloy nanoparticles within these reactors, thereby allowing one to study their interactions with each other and the surrounding solvent. Through *in situ* electron microscopy, we identify that particle pairs at the receding polymer edges coalesce following either conventional steady-state or accelerated kinetics depending on the relative orientation of the nanoparticle pair in relation to the polymer recession direction. Moreover, qualitative analysis of the forces on the nanoparticles within the reactor suggests that an evaporation-induced capillary dragging effect plays a significant role in the accelerated coalescence process.

RESULTS AND DISCUSSION

**Geometry of Polymer Nanoreactors on Amorphous Carbon Films**

Polymer inks prepared by dissolving metal ion precursors, chloroauric acid ($HAuCl_4$) and chloroplatinic acid ($H_2PtCl_6$), in an aqueous solution of PEO-*b*-P2VP were cast onto DPN tips and then deposited onto amorphous carbon films in the form of nanoreactors. Amorphous carbon films are favorable for *in situ* TEM compared to hydrophobic silicon nitride thin films (conventionally used in SPBCL experiments), due to their decreased electron scattering and increased stability under an electron beam.[46-47] The nanoreactors were characterized by HAADF STEM imaging and tapping mode atomic force microscopy (AFM) where a domed morphology is observed (Figure 1A–B). To reduce the vibration of the thin films during AFM imaging, the amorphous carbon film was transferred to a silicon wafer (Supporting Information Text S1 and Figure S1A–B). The polymer nanoreactors have a well-defined, uniform dome shape with a diameter-to-height aspect ratio of ~9:1 (Figure 1C). The average contact angle of the polymer nanoreactors on the carbon film was determined by imaging the polymer domes with AFM and overlapping the height profiles of the edge of multiple domes (Figure S2). By calculating the tangent angle at the edge, we determined a contact angle of 13.4 ± 0.7° (Figure 1D), which is close to that measured optically (12–17°, Figure S3A) and significantly smaller than that in conventional SPBCL experiments on hydrophobic substrates.[41] As a result, polymer sections with a thickness of less than 10 nm are accessible over a large annular region (~50 nm wide) near the nanoreactor edges. For *in situ* TEM experiments, we treat these areas as solvent films that are thin enough to carry nanoparticles with minimal vertical overlap.

**Au-Pt Nanoparticles Generated by Electron Irradiation in Polymer Nanoreactors**

Au-Pt nanoparticles dispersed in polymer nanoreactors were generated *in situ* by locally irradiating the precursor-containing polymer nanoreactors with the electron beam of a TEM. To slow down the kinetics in non-irradiated regions, the sample was cooled to cryogenic conditions (typically < -160 °C) with liquid nitrogen prior to exposure. Upon irradiation, ultrafine nanoparticles with diameters typically between 1–3 nm were rapidly generated within the field of view of a few frames (Figure S4; Text S2). Polymer inks with three different molar ratios of metal ion precursors ($HAuCl_4$:$H_2PtCl_6$ = 30:70, 50:50, 70:30) were used. The elemental distribution of Au and Pt is generally uniform within the electron-irradiated areas where multiple nanoparticles formed (Figure S4). This confirms a direct relationship between precursor and product compositions in the nanoreactors (Figure 2A, green lines; Figure S5). EDS quantification of randomly selected nanoparticles from different nanoreactors shows that the composition of individual nanoparticles is close to (within 6.5 atomic % of Au) the overall polymer nanoreactor composition, which indicates that the particles are uniform and do not form different phases with distinct compositions (Figure 2A, color bars). This observation is in contrast with the large miscibility gap observed for Au-Pt in the bulk which favors phase separation;[18, 48] however, previous reports have shown that homogeneous alloy nanoparticles can be generated from wet chemical syntheses.[14, 16-17, 49]

Additionally, the synthesis of uniform Au-Pt nanoparticles was confirmed by high-resolution transmission electron microscopy (HRTEM), which shows clear lattice structures with no grain boundaries (Figure 2B inset, Figure S6A–C). The size distribution of the nanoparticles is similar for the three compositions (Figure S6D). The {111} reflections of the face-centered cubic (*fcc*) Au-Pt alloy are identified in the Fast Fourier Transform (FFT) radial intensity profile of each HRTEM image (Figure 2B). According to the radial location of these peaks in reciprocal space, the average lattice spacing of the Au-Pt nanoparticles increases from 0.227 to 0.231 to 0.235 nm (peak positions determined by Gaussian fitting) as the Au atomic percentage increases from 30 to 50 and then 70%. The experimentally observed values vary between the {111} lattice spacing of pure Pt (0.227 nm, JCPDS No. 01-070-2057) and pure Au (0.236 nm, JCPDS No. 01-089-3697)[50] as the Au fraction increases, which supports the conclusion that homogeneous alloys were generated.

**Polymer Evaporation Induced by Electron Irradiation**

High energy electrons can induce chain breaking and polymer removal in vacuum at low temperature by both knock-on (electron-induced bond breaking and atom displacement) and thermal effects, known as the electron beam evaporation process.[51-53] While the physical removal of the polymer is visible in the electron microscopy experiment, chain breaking is not observable but may decrease the viscosity of the polymer reactor. To understand the pathways and kinetics of electron-induced polymer evaporation at the nanoscale under the experimental conditions explored herein, polymer nanoreactors containing pre-formed Au nanoparticles as position markers for electron microscopy were prepared. The polymers were scanned with the electron beam to induce evaporation (Movie S1 and Figure 3A) and a region near an Au nanoparticle was monitored (Figure S7A–C) where the nanoparticles exhibit high contrast in the HAADF images (white regions in Figure 3A). Compared to the size of the irradiated region (edge length: 51.79 nm), the aggregate of nanoparticles (~100 nm in size) does not show noticeable movement with respect to the nanoreactor facilitating *in situ* drift compensation. In addition, this nanoparticle aggregate is located deeper in the polymer reactor than the removed polymer layers, since no secondary electron (SE) features are observed in the irradiated area, although the Au nanoparticles are visible in BF and HAADF images (Figure S7D–F). Therefore, the region of study is representative of the polymers and not significantly affected by particles embedded within it.

By analyzing the HAADF movie of sequential scans, where the signal intensity quantified the projected mass for each pixel,[54-56] it was revealed that the polymer is non-uniformly removed by electron irradiation at the nanoscale. At the beginning of the movie, the width of the polymer features gradually shrank, and voids were generated with lower signal intensity (Figure 3A, first row; and Figure 3B, stage 1) creating thin (~10 nm wide) polymer filaments that were then rapidly removed (Figure 3A, middle row; and Figure 3B, stage 2). In the final stage, the projected mass of the whole polymer region gradually decreased (Figure 3A, bottom row; and Figure 3B, stage 3). We also note that the whole polymer region, including the shrinking filaments, shows a variation in intensity on the order of a few nm suggesting that local polymer evaporation plays a significant role in the removal process. HAADF quantification of the polymer region shows a three-stage process which can be aligned with the three stages of filament formation and removal (Figure 3C). Stages 1 and 3 show similar rates of mass loss, which can be attributed to relatively uniform evaporation of polymer. In stage 2, the rate of mass loss is almost an order of magnitude greater suggesting that the kinetics of filament removal are faster. This rapid removal possibly stems from the large surface area of the thin filaments. Note that in this *in situ* STEM experiment, the electron flux in each pixel is the same, and no correlation can be found between the beam scanning direction and polymer evaporation dynamics. Therefore, nanoscale polymer evaporation by the electron beam is both spatially and temporally non-uniform.

**Dynamics of Au-Pt Nanoparticles in Evaporating Polymer Thin Films**

The dynamics of ultrafine Au-Pt nanoparticles in polymer thin films were studied for the 50% Au case by recording sequential ADF images (Movie S2, Figure 4A) close to the edge of a polymer nanoreactor. In the ADF images, the Au-Pt nanoparticles (bright contrast) are dispersed in the polymer solvent (gray contrast against the dark background). This polymer region evaporates non-uniformly during irradiation.

To study the interaction between the polymers and nanoparticles, the nanoparticles in each frame were tracked (Figure 4A). The movement of nanoparticles close to the polymer edge is associated with the retraction of polymers, as evidenced by the correlation between the receding direction of the polymer edges and the movement of neighboring nanoparticles. This observation suggests that the dynamics of the polymer film resembles the flowing and drying of a thin, viscous fluid film on a surface. Importantly, the shrinking polymer filaments drag the nanoparticles along a converging line (Red dashed line **L1** labeled in Figure 4A, Scan 25) as seen by the decrease in the distance between the nanoparticles and **L1** (Figure 4B). As a result, the nanoparticles form a network-like distribution on the surface after the polymers are fully removed, leaving areas where there are almost no nanoparticles.

During the evaporation process, the nanoparticles undergo substantial coalescence as seen by the decrease in their number (Figure 4C). In a representative 25 frame study, the number of nanoparticles decreased from 106 to 49 through 55 merging and 12 new particle formation events. Therefore, coalescence contributes to nearly 80% of the particle number reduction.

In addition to coalescence, other types of nanoparticle dynamics typical in solutions were observed. For instance, small nanoparticles/clusters, < 1 nm, re-dissolve into the polymer solvent, especially when close to larger nanoparticles (Figure S8A). This observation is typical of Ostwald ripening in which the chemical potential gradient drives the mass transport of atomic constituents from smaller particles to larger ones. In addition, the removal of the polymer solvent results in local supersaturation of dissolved metal species, which supplies atomic constituents either for the growth of existing nanoparticles (Figure S8B) or the nucleation of new particles (Figure S8C).

**Coalescence Kinetics of Au-Pt Nanoparticles at Receding Polymer Edges**

The kinetics of particle coalescence were monitored over four sequential frames before a nanoparticle pair fully coalesced into a single particle (Figure 5A and C). Two types of coalescence kinetics are observed. In the first type, the nanoparticles gradually migrate towards each other as in conventional bulk solutions (Figure 5A and Figure S9A).[57] This occurred either when there is enough polymer around the nanoparticle pair (Figure S9A) such that they do not interact with the polymer dome boundary, or when they both reside along the polymer edge (Figure 5A). Quantitatively, the edge-to-edge distance between all of these nanoparticle pairs decreases in a linear manner, which suggests that the migration kinetics of these nanoparticles are in a steady state (Figure 5B).

In the second type, the coalescence shows a sudden acceleration. This occurs when one of the particles interacts with the edge of the receding polymers, which points one particle towards another (Figure 5C and Figure S9B). The distortion of nanoparticles observed in scan # -1 reveals that these particles move at a faster rate than the time needed to acquire a single frame. Given the pixel dwell time of 7.5 µs, such movements occur on a timescale of tens to hundreds of µs. The edge-to-edge distance of the nanoparticle pairs undergoing this coalescence mode does not decrease significantly before the accelerated movement. On the contrary, linear fits predict that no coalescence should occur at scan # 0 (Figure 5D). The kinetics in this mode are different from conventional, steady-state coalescence, and largely driven by the dynamics of the evaporating polymer. In particular, capillary forces may contribute to the acceleration since the polymer edges are interacting with a nanoparticle and moving in the direction of the other particle.

**Force States of the Coalescing Nanoparticles**

The observation that nanoparticle pairs approach each other at a steady rate (Figure 5B), when not interacting with a receding edge, suggests that these particles are under a *quasi*-equilibrium force state. In other words, the total force on each nanoparticle along the line connecting them is close to zero before the particles are in direct contact. This is also evidenced by the Brownian motion of nanoparticles in polymers when evaporation is not a dominant factor (Text S3 and Movie S3). On the other hand, nanoparticles at receding polymer edges experience an unbalanced total force. When the receding edge is perpendicular to the line connecting a nanoparticle pair, the capillary force exerted on the particle at the edge is not balanced, and this accelerates nanoparticle coalescence. The capillary effect was qualitatively analyzed using a planar model, where the edge of a thin, continuous fluid film with thickness, $\delta$, is passing through a fixed circular rigid disk of radius, $R$, in the $z$ direction (Figure 6A). The contact angle of the polymer on gold, $\theta$, is small according to the optical measurement (Figure S3B). The fluid edges at a distance larger than $L$ from the particle are kept horizontal (red lines) and do not interact with the particle. The polymer edge $< L$ from the particle forms a meniscus due to contact with the particle and is approximated as a circular arc since gravity is negligible (Bo number $\ll 1$). A dimensionless plot of the total capillary force on the particle in the $z$ direction as a function of fluid edge position at semi-static conditions (Figure 6B) is generated as detailed in the Supporting Information (Text S4). By varying the dimensionless length, $l = L/R$, and $\theta$, we find that the total force increases as the fluid edge sweeps across the majority of the particle. When the fluid edge reaches the middle of the particle ($H/R \approx 1$), which is frequently observed in the TEM experiments (Figure 5C), the vertical force reaches $F_z \sim \gamma \delta$. Note, in three dimensions, an additional interface between the fluid and the top surface of the particle will also contribute a capillary force in the same direction with similar magnitude. Given that the typical surface energy of PEO is ~40 mN/m,[58] the drag force exerted on a 1 nm particle is ~$10^{-1}$ nN, which is comparable to or larger than the typical solvation and van der Waals interactions between nanoparticles of similar sizes.[59-60] Therefore, the fluid edge dragging effect, which increases as the fluid edge passes through the particle, is able to induce the acceleration of the particle before coalescence.

CONCLUSIONS

In summary, we have shown how polymer nanoreactors, when coupled with *in situ* electron microscopy, can be used to reveal the evaporation dynamics of PEO-*b*-P2VP polymers, and the interaction between thin films and ultrafine Au-Pt nanoparticles under electron irradiation. This work has two important implications for understanding the interactions between particles and solvents. First, nanostructures containing elements with a large miscibility gap at room temperature (e.g., Au-Pt) may be generated by coalescence of ultrafine alloy nanoparticles, which can be first synthesized by fast reduction with a strong reducing source (e.g., an electron beam in this study). Such kinetically trapped nanostructures may result in the formation of relatively stable particles without phase separation, as reported in the literature.[14, 16-17, 49] Second, solvent evaporation can dramatically alter the kinetics of nanoparticle coarsening and particle spatial distribution on surfaces. Consequently, conventional models for calculating coalescence and ripening rates may not directly apply to such surface systems.

Collectively, these conclusions provide important insight into the SPBCL process, in particular how smaller particles at early stages of the process coarsen to form larger multicomponent nanoparticles.[61] Since polymer decomposition in the SPBCL process is likely non-uniform, as evidenced by the spatial distribution of the carbon residue (Figure S10), polymer decomposition during the nanoparticle coarsening stage likely accelerates the *local* aggregation of small nanoparticles into intermediate sized ones. However, this can impede further coarsening of the remaining particles (because there will not be polymer forcing the intermediate sized particles together) and the targeted outcome of one nanoparticle per nanoreactor. Note that a large library of metal ion precursors can be used in polymer nanoreactors,[42-43] enabling the study of interactions in even more complicated, multicomponent nanoparticle systems. Moreover, other non-volatile solvent systems such as ionic liquids,[38-39] can also be potentially explored using the nanoreactor approach. Therefore, this strategy should provide a means for garnering significant insight into the dynamics and interactions of complex nanoparticles in fluidic dispersions.

EXPERIMENTAL SECTION

**Materials.** Chloroauric acid hydrate (HAuCl$_4$·xH$_2$O, ≥99.9% trace metals basis) and chloroplatinic acid hydrate (H$_2$PtCl$_6$·xH$_2$O, ≥99.9% trace metals basis) were purchased from Sigma-Aldrich and used without further purification. Poly(ethylene oxide)-*block*-poly(2-vinylpyridine) (PEO-*b*-P2VP, $M_n$ = 2.8-*b*-1.5 kg/mol, polydispersity index = 1.11) was purchased from Polymer Source, Inc. Dip-pen nanolithography (DPN) 1D pen arrays (type M, with gold coating) were purchased from Nanoink, Inc. ScanAsyst-Air atomic force microscopy (AFM) probes were purchased from Bruker Nano Inc. Amorphous carbon films supported on copper TEM grids were purchased from Ted Pella Inc. and Electron Microscopy Sciences. Deionized water was produced from a Millipore Milli-Q purification system (resistivity at 25 °C = 18.2 MΩ·cm).

**Precursor-containing Polymer Ink Preparation.** PEO-*b*-P2VP and metal ion precursors (HAuCl$_4$ and H$_2$PtCl$_6$, respectively) were dissolved in deionized water with deliberately varied molar ratios of metal species. The resulting solution had a polymer concentration of 5 mg/mL and a 4:1 molar ratio of pyridyl groups to total metal species. Freshly prepared inks were stirred for at least 4 h at room temperature prior to use.

**Au Nanoparticle-containing Polymer Ink Preparation.** PEO-*b*-P2VP polymer inks containing HAuCl$_4$ were prepared following the methods described above. This ink was stirred at 60 °C for at least two weeks to ensure the complete formation of Au nanoparticles *in situ*. Complete particle formation was established when the characteristic red color, which indicates the formation of Au nanoparticles, did not change further. No other metal species or surfactants were added to the ink.

**Patterning of Nanoreactors.** DPN pen arrays were coated with inks and dried under ambient conditions. After the inking process, the pen arrays were mounted onto an AFM (Park Systems XE-150) in a chamber with a temperature of 20 °C and relative humidity of 78%. Polymer nanoreactors were deposited by bringing the pen arrays in contact with the amorphous carbon films.

**Atomic Force Microscopy (AFM).** Amorphous carbon films were transferred to Si wafers prior to AFM measurements (See Supporting Information Text S1). AFM measurements were performed on a Bruker Dimension Icon in tapping mode using Bruker ScanAsyst-Air AFM probes (k = 0.4 N/m).

**Cryogenic Electron Microscopy.** A Hitachi HD-2300A equipped with Thermo Scientific NORAN System 7 energy dispersive x-ray spectroscopy (EDS) detectors was operated at 200 kV for scanning transmission electron microscopy (STEM) imaging in annular dark field (ADF), high-angle ADF (HAADF), bright field (BF) and secondary electron (SE) modes, as well as EDS analysis. A JEOL JEM-3200FS was operated at 300 kV for high-resolution transmission electron microscopy (HRTEM). In both microscopes, Gatan 626 single tilt liquid nitrogen cryo transfer holders were used to support the samples. The samples were inserted into the microscope column, and then cooled down to cryogenic conditions under vacuum by liquid nitrogen before electron irradiation. *In situ* (HA)ADF movies were recorded at a resolution of 512 px × 512 px, scan area of 51.79 nm × 51.79 nm, and pixel dwell time of 20 µs (Movie S1) or 7.5 µs (Movie S2). The probe current was 0.168 nA. EDS analysis was performed after fixing the area of interest by scanning the area using a lower flux electron beam until the nanoparticles did not grow further. EDS of individual nanoparticles was performed with manual drift correction based on the HAADF signal. EDS quantification was performed with the Thermo Scientific NSS software after background removal and matrix correction by the Proza method (Phi-Rho-Z). A Gatan K2 Summit direct electron detector was used for HRTEM imaging at a resolution of 3710 px × 3838 px (pixel size is 0.013 nm for the imaging conditions used).

**Lattice Spacing Analysis.** The scale of HRTEM images was calibrated based on a standard Au (100) thin film sample supplied by Ted Pella Inc. Fast Fourier Transform (FFT) was performed on HRTEM images, which generated a 4096 px × 4096 px image of the real part. The spatial frequencies per pixel in the FFT image in x and y directions are equal. Average radial intensity profiles were calculated by integrating the central circular area of the FFT image with a radius of 512 px. The background due to the amorphous carbon substrate and polymers was fitted and subtracted in the final presentation in Figure 2B.

**Noise Reduction.** We estimated the standard deviation of the noise by filtering the Fourier transform of a frame from the movies with a long-pass filter cut off at 1/5 px$^{-1}$, then inverse transformed the result back to the real space, and finally calculated the intensity standard deviation, $\sigma_N$, of this processed image in the real space. The movies were subsequently processed with a modified code using the video block-matching and 3D filtering (V-BM3D) method with the estimated noise standard deviation $\sigma_N$.[62] Intensity declipping was skipped during the denoising process.

**Particle Tracking.** Particles in the denoised movies were analyzed using the TrackMate package in ImageJ (available for free through the NIH).[63] Particle locations were detected by a Laplacian of Gaussian (LoG) filter detector assisted by manual corrections. Particle movement was linked between each frame using a Linear Assignment Problem (LAP) tracker with the maximum linking/gap-closing/splitting/merging distances set at 30 px, or 3.0 nm.

## ASSOCIATED CONTENT

**Supporting Information**. Details and additional discussions on transferring carbon films, imaging conditions, and nanoparticle force states; additional AFM and electron microscopy results; and movies S1−S2. This material is available free of charge via the Internet at http://pubs.acs.org.

## AUTHOR INFORMATION

**Corresponding Author**

*chadnano@northwestern.edu (C.A.M.), v-dravid@northwestern.edu (V.P.D.)

**Present Address**

⊥School of Materials Science and Engineering, Sun Yat-sen University, Guangzhou, Guangdong 510006, China (Z.X.)

**Notes**

The authors declare no competing financial interest.

ACKNOWLEDGMENT

We thank Prof. Monica Olvera de la Cruz, Dr. Mykola Tasinkevych, Mr. Yaohua Li, and Dr. Sara M. Rupich (Northwestern University, NU) for helpful discussions, and Dr. Christos D. Malliakas (NU) for help with thermal analysis and mass spectrometry. This material is based upon work supported by the Sherman Fairchild Foundation, Inc., GlaxoSmithKline LLC, the Air Force Office of Scientific Research awards FA9550-12-1-0280 and FA9550-16-1-0150, and the Air Force Research Laboratory under agreement number FA8650-15-2-5518.  The U.S. Government is authorized to reproduce and distribute reprints for Governmental purposes notwithstanding any copyright notation thereon.  The views and conclusions contained herein are those of the authors and should not be interpreted as necessarily representing the official policies or endorsements, either expressed or implied, of Air Force Research Laboratory or the U.S. Government. B.M. acknowledges support from the Eden and Steven Romick Post-Doctoral Fellowship through the American Committee for the Weizmann Institute of Science. This work made use of the EPIC Facilities of the NUANCE Center supported by the SHyNE Resource NNCI site (NSF ECCS-1542205), the MRSEC program (NSF DMR-1720139), the IIN, and the State of Illinois through the IIN; and the Structural Biology Facility supported by NCI CCSG P30 CA060553 awarded to the Robert H Lurie Comprehensive Cancer Center, and the Chicago Biomedical Consortium with support from the Searle Funds at The Chicago Community Trust.

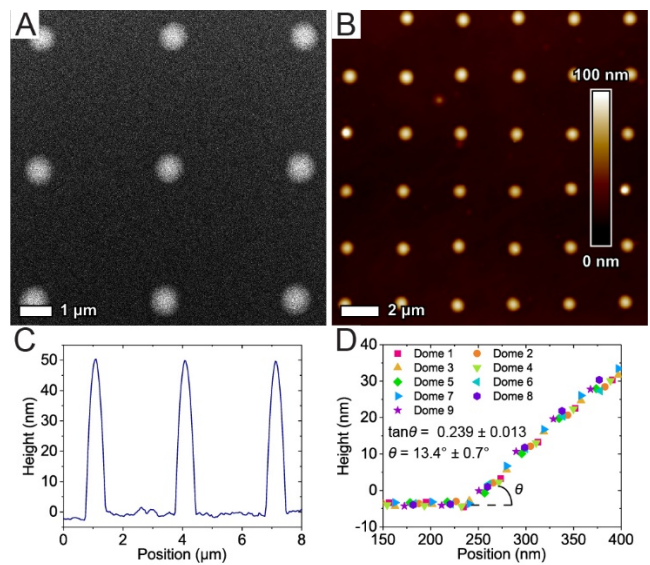

**Figure 1.** Polymer nanoreactors on amorphous carbon films. (A) HAADF, and (B) AFM height images of a nanoreactor array. (C) AFM height profile of three nanoreactor domes across their centers. (D) Plot showing the extrapolated contact angle from nine AFM height profiles at the edge of the nanoreactor domes.

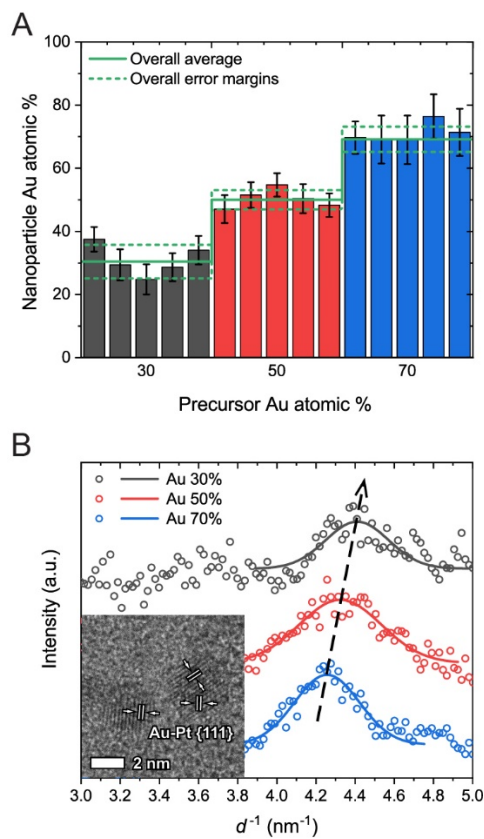

**Figure 2.** Au-Pt nanoparticles formed by local electron irradiation in polymer nanoreactors. (A) EDS quantification of the Au atomic percentage in polymer nanoreactors with different precursor compositions. The bars represent randomly chosen nanoparticles at a specific precursor composition (black: 30% Au, red: 50% and blue: 70%), and the green lines represent the overall Au atomic percentages in each nanoreactor. Error bars and margins represent EDS quantification uncertainties. (B) FFT radial intensity profiles (circles) and corresponding Gaussian fits (curves) as a function of reciprocal spacing for regions near the Au and Pt {111} reflection. Inset: A typical HRTEM image of Au-Pt nanoparticles showing their {111} lattice fringes.

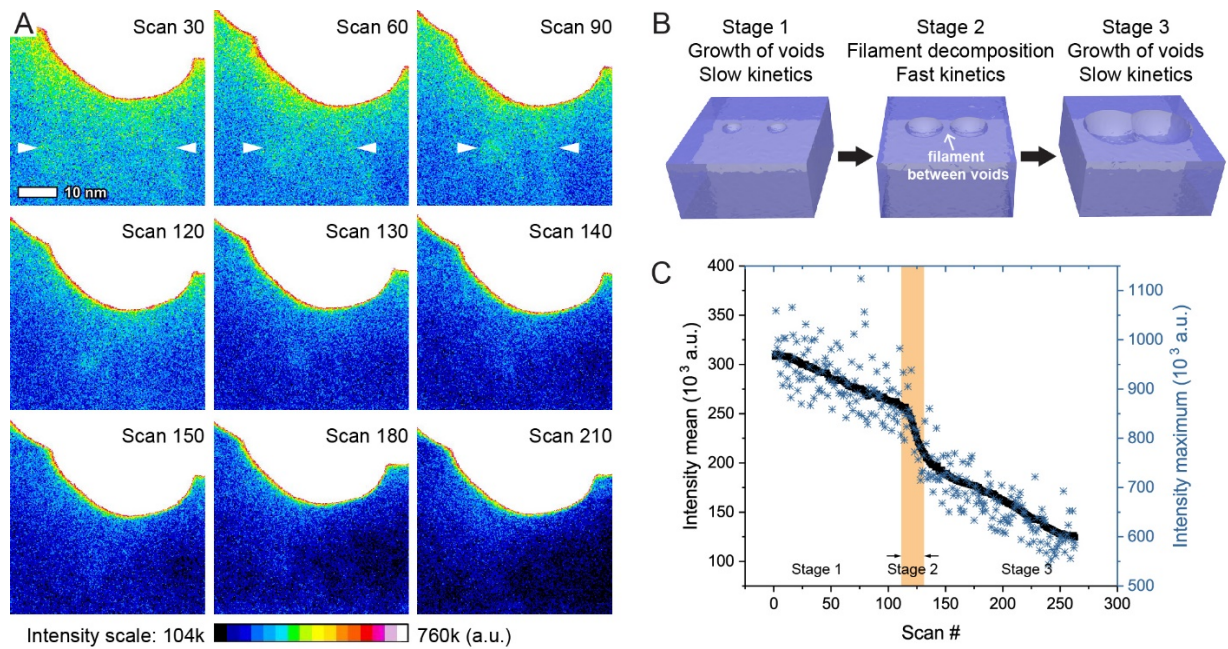

**Figure 3.** Non-uniform evaporation of polymers via electron irradiation. (A) False-colored HAADF images showing the removal of polymer as a function of time (i.e., scan number). The white regions on the top of the images are Au nanoparticles (20–30 nm diameter) that serve as position markers for drift compensation. Pairs of white arrows indicate the shrinkage of the polymer filaments. Below: intensity color bar. (B) Schematic of the three-stage polymer evaporation process involving slow growth of voids and fast removal of filaments between voids. (C) HAADF signal quantification of the mean (black) and maximum (blue) pixel intensity for the polymer region undergoing evaporation shows multiple stages.

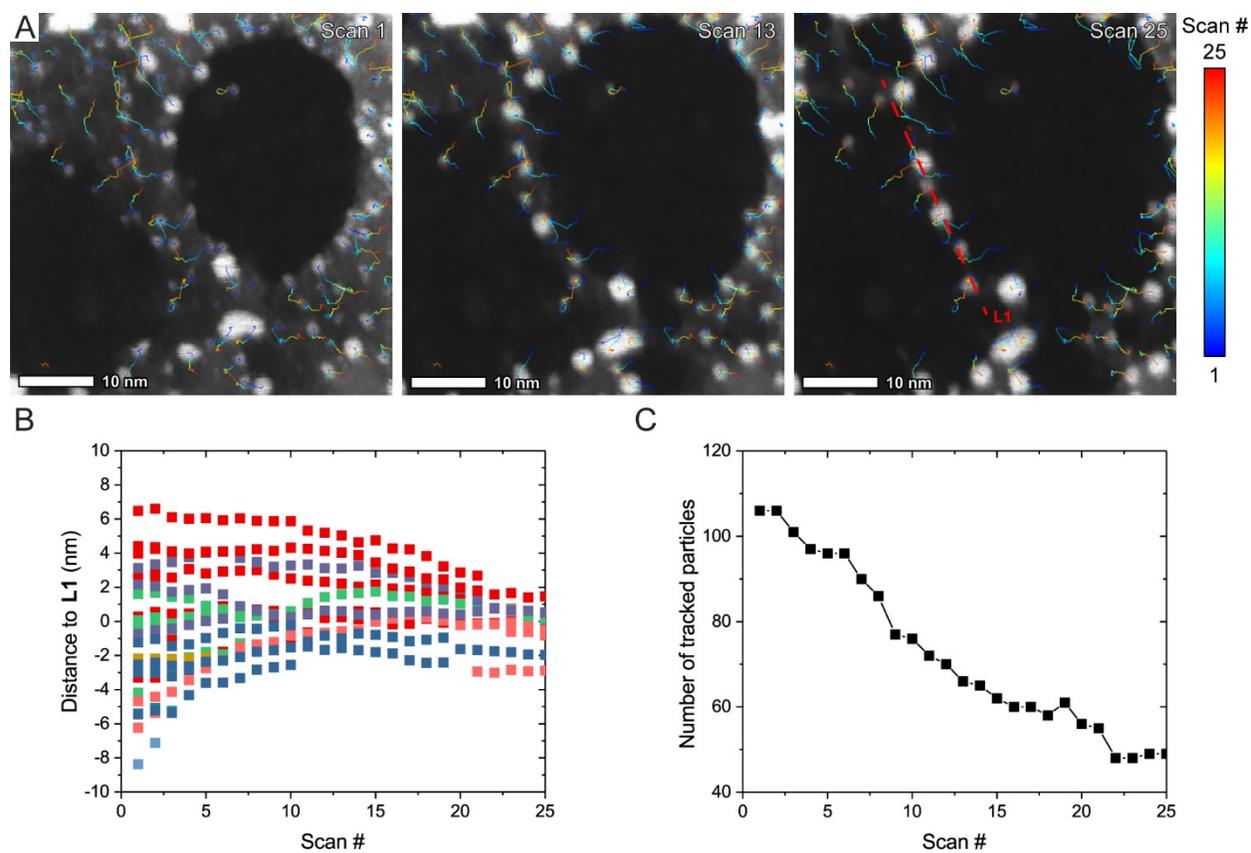

**Figure 4.** Migration of Au-Pt nanoparticles induced by non-uniform polymer evaporation. (A) Denoised ADF images and tracking results of Au-Pt nanoparticles as a function of scan number. (B) Distance from nanoparticles in the filament region to the filament center [**L1** in (A), red dashes] as a function of scan number. Data points of the same color represent nanoparticles that coalesce over the course of the experiment, and therefore, are considered to share the same track. Nanoparticles to the right of **L1** are defined as having positive distances while those on the left have negative ones. (C) Number of tracked nanoparticles as a function of scan number.

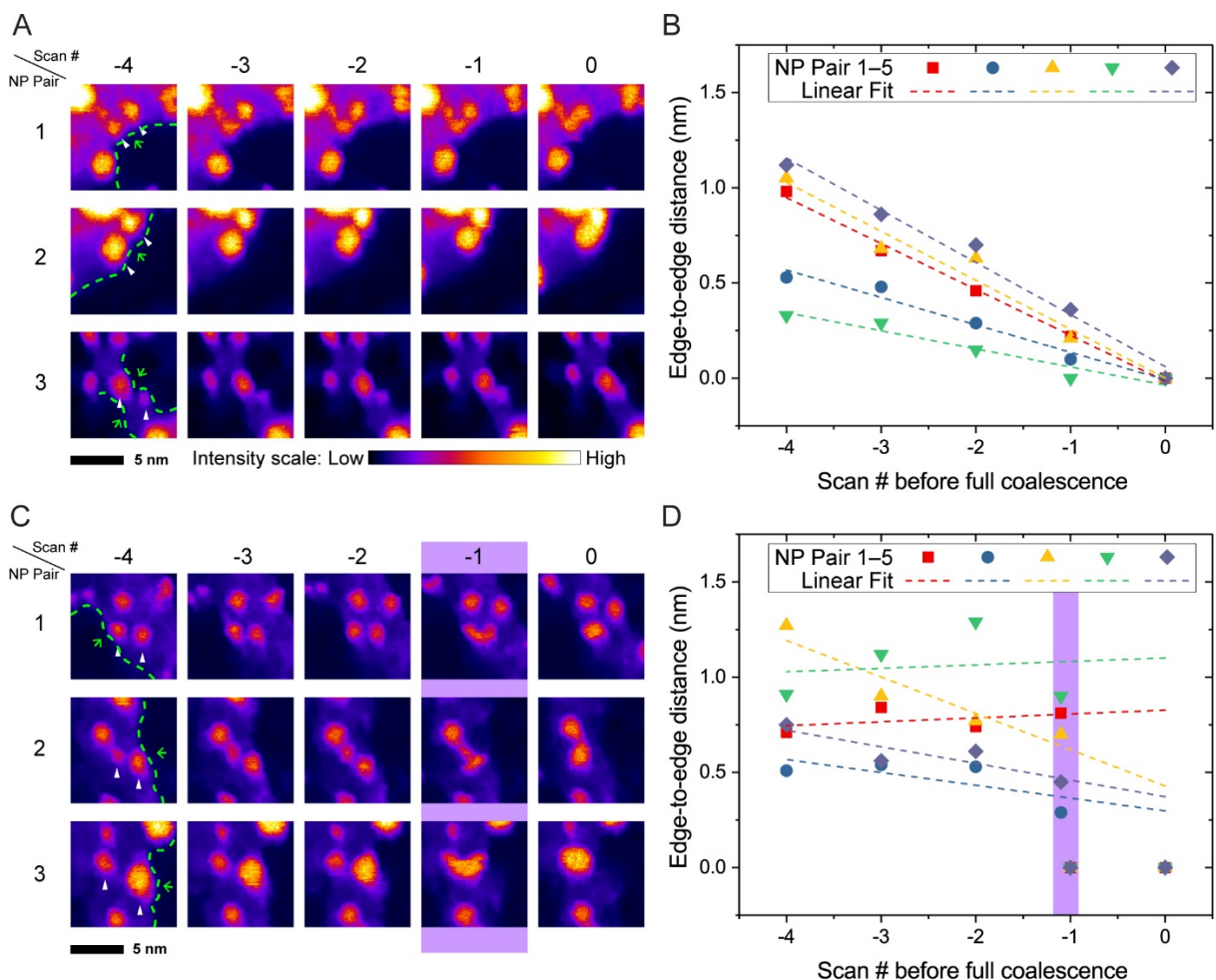

**Figure 5.** Coalescence of Au-Pt nanoparticles following conventional kinetics (A, B) or undergoing acceleration before contact (C, D). (A, C) Sequential color-enhanced, denoised ADF images showing the coalescence between nanoparticle pairs (NP Pairs), indicated by white arrows. Green dashes and arrows show the edge of the polymer medium. Scan # indicates the relative frame number before the NP pairs fully coalesce. An abrupt location change of nanoparticles is observed at Scan # -1 in (C) (purple box). All images have the same scale bar. (B, D) Edge-to-edge distances and linear fits of each NP pair as a function of scan number before coalescence. The first data points at Scan # -1 in (D) are inferred from the location before the abrupt location change. The linear fits in (D) are based on Scan # -4 to -1 (first data point).

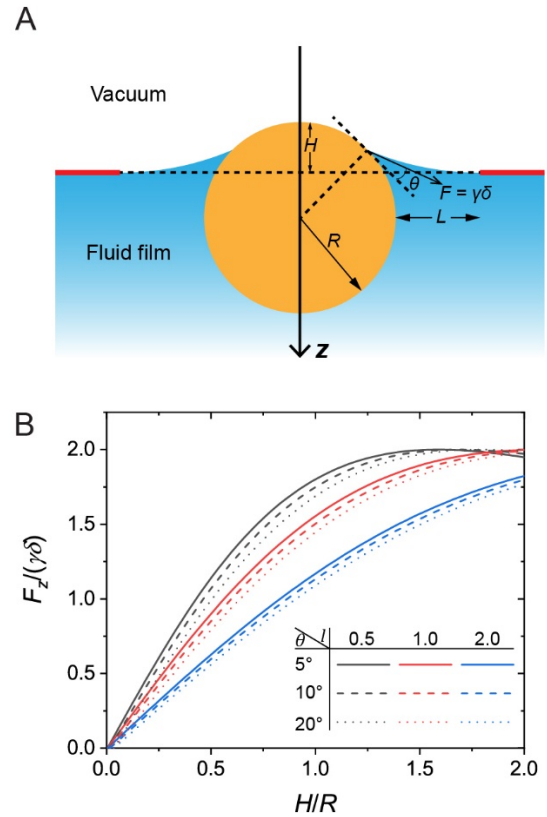

**Figure 6.** Qualitative force state analysis of a particle at a receding fluid edge. (A) Top down schematic of a particle at the fluid edge with the model parameters indicated. The fluid edge at a distance greater than $L$ is kept horizontal (red lines). The **z** direction is the fluid receding direction. (B) Dimensionless plot of the total capillary force on a particle in the **z** direction as a function of the relative location of the fluid edge in relation to the total particle size. Nine scenarios based on different dimensionless length, $l = L/R$, and contact angle, $\theta$, are shown.

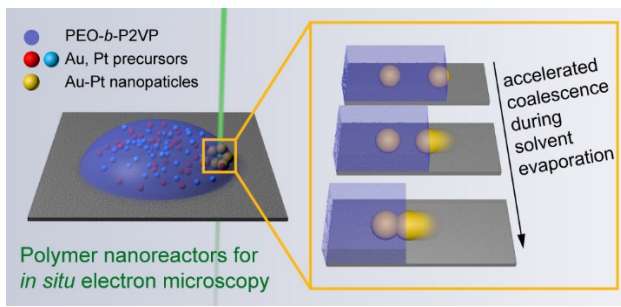